\documentclass[11pt]{article}
\usepackage{graphicx}

\begin{document} 

\def\xslash#1{{\rlap{$#1$}/}}
\def \p {\partial}
\def \dd {\psi_{u\bar dg}}
\def \ddp {\psi_{u\bar dgg}}
\def \pq {\psi_{u\bar d\bar uu}}
\def \jpsi {J/\psi}
\def \psip {\psi^\prime}
\def \to {\rightarrow}
\def \lrto{\leftrightarrow} 
\def\bfsig{\mbox{\boldmath$\sigma$}}
\def\DT{\mbox{\boldmath$\Delta_T $}}
\def\xit{\mbox{\boldmath$\xi_\perp $}}
\def \jpsi {J/\psi}
\def\bfej{\mbox{\boldmath$\varepsilon$}}
\def \t {\tilde}
\def\epn {\varepsilon}
\def \up {\uparrow}
\def \dn {\downarrow}
\def \da {\dagger}
\def \pn3 {\phi_{u\bar d g}}

\def \p4n {\phi_{u\bar d gg}}

\def \bx {\bar x}
\def \by {\bar y}
\begin{center} 
{\Large\bf   Twist-3 double-spin asymmetries in Drell-Yan processes }
\par\vskip20pt
M.C. Hu$^{1,2}$,  J.P. Ma$^{1,2,3}$,  Z.Y. Pang$^{1,2}$ and G.P. Zhang$^{4}$    \\
{\small {\it
$^1$ CAS Key Laboratory of Theoretical Physics, Institute of Theoretical Physics, P.O. Box 2735, Chinese Academy of Sciences, Beijing 100190, China\\
$^2$ School of Physical Sciences, University of Chinese Academy of Sciences, Beijing 100049, China\\
$^3$ School of Physics and Center for High-Energy Physics, Peking University, Beijing 100871, China\\
$^4$ Department of Physics, Yunnan University, Kunming, Yunnan 650091, China}} \\
\end{center}


\vskip 1cm
\begin{abstract}
We study double-spin asymmetries in Drell-Yan processes in which one initial hadron is transversely polarized and another one is longitudinally polarized. The complete part of the hadronic tensor relevant to asymmetries is derived. This part consists of twist-2 and twist-3 parton distributions and is gauge invariant. We construct some observables which can be used to extract these parton distributions from experimental measurements.

 \end{abstract}      
\vskip 5mm

\vskip40pt

\noindent 
{\bf 1. INTRODUCTION}

\par\vskip5pt 

Predictions about high energy scattering of hadrons of large momentum transfers can be made 
with QCD factorizations. At the leading power of the inverse of large momentum transfers, cross sections 
can be predicted with collinear twist-2 parton distributions convoluted with perturbative coefficient functions. These parton distributions contain information about inner structures of hadrons and are nonperturbative. Currently, 
the twist-2 parton distributions are well known and used to make predictions of various processes. 

Beyond the leading power,  the contributions to cross sections are factorized with twist-3 or higher twist parton distributions. Although they are power suppressed, but with the progress of the experiment it is now 
possible to measure them. An example is of single transverse-spin asymmetries, which have been observed in early experiments in\cite{E704,hermes}.  Such asymmetries are factorized with twist-3 parton distributions as pointed out in \cite{QiuSt,EFTE}.  
The importance for studying and measuring twist-3 parton distributions is that they contain information about hadron's inner structure more than twist-2 parton distributions. 

In this work,  we study with QCD collinear factorization double-spin asymmetries in Drell-Yan processes where one initial hadron is transversely polarized and another is longitudinally polarized. These double-spin asymmetries at the leading power can be factorized with collinear twist-3 parton distributions.  Measurements of these asymmetries will help to learn twist-3 parton distributions. The double-spin asymmetries in Drell-Yan processes have been studied in \cite{JaJi, BMT,KTS} at leading order of $\alpha_s$.  At the order, the transverse momentum $q_\perp$ of the lepton pair is small, or approximately zero, and because of that there is no hard parton radiation. One may expect that the relevant part of the hadronic tensor 
is proportional to $\delta^2 (q_\perp)$. This is true for the twist-2 part. However, at twist-3 the hadronic tensor 
contains not only a part with $\delta^2 (q_\perp)$, but also a part with the derivative of $\delta^2 (q_\perp)$, as shown in \cite{MZ1,CMZ}. Each part alone is not electromagnetic gauge invariant. Only the sum of the two parts is gauge invariant.  In this work, we will derive the complete part of the hadronic tensor for the double-spin asymmetries. With our result more double-spin asymmetries can be predicted with twist-3 parton distributions.  

At the next-to-leading order of $\alpha_s$, the transverse momentum can be large. The contribution from the next-to-leading order is the one-loop correction  to the tree-level results of the proposed observables here.  It is noted that the correction can contain collinear divergences. It is expected that such collinear divergences can be factorized into various parton distributions, as shown in an explicit calculation of one-loop correction to single transverse-spin asymmetries at twist-3 of Drell-Yan processes in \cite{CMZ2}.  
 
 Our work is organized as the following: In Sect.II we introduce our notations and definitions of 
collinear  twist-3 parton distributions. Relations among these distributions are discussed. In Sect.III we derive 
 the hadronic tensor for double-spin asymmetries, where we explain in detail how the contribution with the derivative of $\delta^2(q_\perp)$ arises. In Sect.IV we construct observables of double-spin asymmetries. Sect.V is our summary.

\par

\par\vskip20pt
\noindent 
{\bf 2. NOTATIONS AND PARTON DISTRIBUTIONS}

\par\vskip5pt 
We consider the Drell-Yan processes:
\begin{equation}
  h_A ( P_A, s_\perp) + h_B(P_B, \lambda_B) \to \gamma^* (q) +X \to  \ell^-(k_1)  + \ell ^+(k_2)  + X,
\end{equation}
where $h_{A,B}$ is a spin-1/2 hadron.
To study the process it is convenient to use the  light-cone coordinate system, in which a
vector $a^\mu$ is expressed as $a^\mu = (a^+, a^-, \vec a_\perp) =
((a^0+a^3)/\sqrt{2}, (a^0-a^3)/\sqrt{2}, a^1, a^2)$ and $a_\perp^2
=(a^1)^2+(a^2)^2$. In this system we introduce two light-cone vectors 
$l^\mu =(1,0,0,0)$ and $n^\mu = (0,1,0,0)$. With the two vectors one can define 
the metric $g_\perp^{\mu\nu}$ and the totally antisymmetric tensor $\epsilon_\perp^{\mu\nu}$ in the transverse 
space, 
\begin{equation}
  g_\perp^{\mu\nu} = g^{\mu\nu} - n^\mu l^\nu - n^\nu l^\mu,
  \quad 
  \epsilon_\perp^{\mu\nu} =\epsilon^{\alpha\beta\mu\nu}l_\alpha n_\beta, \quad \epsilon_\perp^{12}=-\epsilon_\perp^{21} =1. 
\end{equation}
We take a frame in which the momenta of hadrons are given by 
\begin{equation}
  P_A^\mu \approx (P_A^+,0,0,0), \quad P_B^\mu  \approx (0,P_B^-,0,0).
 \end{equation}
We consider the case that $h_A$ is transversely polarized with the spin vector $s_\perp^\mu =(0,0,s_\perp^1,s_\perp^2)$, 
and $h_B$ is longitudinally polarized with the helicity $\lambda=\pm 1$.  The hadronic tensor 
 $W^{\mu\nu}$  for the process is defined as 
\begin{eqnarray}
  W^{\mu\nu} =  \sum_X \int \frac{d^4 x}{(2\pi)^4} e^{iq \cdot x} \langle h_A, h_B  \vert
    \bar \psi(0) \gamma^\nu \psi(0) \vert X\rangle \langle X \vert \bar \psi (x) \gamma^\mu \psi(x) \vert
     h_B,h_A \rangle,   
\end{eqnarray}
where $q$ is the momentum of the lepton pair. Its invariant mass  is $q^2=Q^2$. 

The hadronic tensor contains all information about the strong interaction in the process.  From the tensor one can obtain the differential cross section,  
\begin{equation} 
  \frac{d\sigma}{d Q^2 d\Omega} = \sum_q  \frac{(4 \pi \alpha e_q)^2}{64\pi^2 S Q^4} \int d^4 q \delta (q^2-Q^2) L_{\mu\nu}W^{\mu\nu} , 
 \label{DSIG} 
\end{equation} 
where $\Omega$ is the solid angle of the lepton in a chosen frame. A commonly used one is  the Collins-Soper frame. Note that $e_q$ is the charge fraction of the quark in unit of $e$. The sum is over flavors of quarks. Besides 
the differential cross section or angular distribution, one can introduce the so-called weighted cross section,
\begin{equation} 
  \frac{d\sigma \langle {\mathcal F} \rangle }{d Q^2 d\Omega} = \sum_q\frac{ (4 \pi \alpha e_q)^2}{64\pi^2 S Q^4} \int d^4 q \delta (q^2-Q^2) L_{\mu\nu}W^{\mu\nu}  {\mathcal F}, 
 \label{WDSIG}  
\end{equation} 
i.e., the event distribution is reweighted by a weight factor ${\mathcal F}$, which can be dependent on lepton momenta and $q_\perp$.  Taking ${\mathcal F}=1$, one obtains the standard differential cross section as given in Eq.(\ref{DSIG}).  In this work we will consider weights which are proportional to $q_\perp$. It is noted that the transverse momentum $q_\perp$ is integrated in Eq.(\ref{DSIG},\ref{WDSIG}).

In the case of $Q\gg \Lambda_{QCD}$ the hadronic tensor can be factorized in QCD collinear factorization. At the leading power of the inverse of $Q$,  
the tensor can be written as convolutions of twist-2 parton distributions with perturbative coefficient functions.    
At this order, the asymmetries appearing, in the case that only one initial hadron is transversely polarized, are 
zero. 
At the next-to-leading power,   the discussed double-spin asymmetries and single transverse-spin asymmetries become nonzero.  The part of the hadronic tensor relevant to these asymmetries can be factorized with twist-3 parton distributions. For our purpose, we discuss in the below the definitions and relations 
of relevant twist-3 parton distributions.  
 
Since we work at the leading order of $\alpha_s$, the relevant parton distributions involve quark fields. 
From the quark density matrix, we can define  twist-2 and twist-3  quark distributions of a hadron  with the momentum $P^\mu \approx ( P^+,0,0,0)$  as follows\cite{JaJi,PDFFF,JKT}: 
\begin{eqnarray} 
 && \int \frac{ d\lambda }{2\pi}e^{-i \lambda  x P^+ } \langle h \vert \left (  \bar \psi (yn) 
  {\mathcal L}_n (yn) \right )_\beta \left ( {\mathcal L}^\dagger  _n (0)  \psi (0) \right )_\alpha  \vert h \rangle
=   \frac{1}{2  } \biggr [ q (x) \gamma^- + \lambda  \Delta q (x)\gamma_5 \gamma^- 
\nonumber\\ 
   && \quad \quad \quad + h_1 (x) \gamma_5 \gamma\cdot s_\perp \gamma^-  \biggr ]_{\alpha\beta}   
  + \frac{1 }{2 P^+} \biggr [ e (x) + q_T (x) \gamma_5 \gamma\cdot s_\perp + \frac{1}{2}\lambda h_L (x) i\sigma^{+-}\gamma_5 \biggr ]_{\alpha\beta} + \cdots, 
\label{TW2PDF} 
\end{eqnarray}
where $\alpha$ and $\beta$ are Dirac indices.  Terms beyond twist-3 are denoted with $\cdots$. They are irrelevant here.  
Note that ${\mathcal L}_n (x)$ is  the gauge link defined as  
\begin{equation} 
    {\mathcal L}_n (x) = P\exp\left \{ -i g_s \int_{-\infty}^0 d\lambda n\cdot G(\lambda n +x) \right \} . 
\label{GLF}                                
\end{equation}
In Eq.(\ref{TW2PDF}) $\lambda$ is the helicity of the hadron.  The transverse spin is given by the vector $s_\perp^\mu$.  In the above, $q(x)$ or $\Delta q(x)$  is the unpolarized or longitudinally polarized quark  distribution, respectively. Note that $h_1(x)$ is the transversity distribution. These distributions are of twist-2. The remaining three distributions are of twist-3.

Besides the twist-3 quark distributions given in the above,  there are other twist-3 parton distributions, 
which can be defined by sandwiching the operator of the gluon field strength tensor, covariant derivative, or derivative in the quark density matrix in 
Eq.(\ref{TW2PDF}). We introduce three types of twist-3 matrix elements as in the following: 
\begin{eqnarray} 
{\mathcal M}^\mu_{F\alpha\beta} (x_1,x_2)  &=&  g_s \int\frac{d \lambda_1 d\lambda_2}{2\pi} e^{-i\lambda_1x_1 P^+ -i \lambda_2 (x_2-x_1) P^+} 
   \langle h \vert \bar \psi_\beta (\lambda_1 n) G^{+\mu}(\lambda_2 n)  \psi_\alpha (0) \vert h \rangle,
\nonumber\\
{\mathcal M}^\mu_{D\alpha\beta} (x_1,x_2)  &=&  P^+  \int\frac{d \lambda_1 d\lambda_2}{2\pi} e^{-i\lambda_1x_1 P^+ -i \lambda_2 (x_2-x_1) P^+} 
   \langle h \vert \bar \psi_\beta (\lambda_1 n) D_\perp ^{\mu}(\lambda_2 n)  \psi_\alpha (0) \vert h \rangle,
 \nonumber\\ 
{\mathcal M}^\mu_{\partial \alpha\beta} (x) &=&  \int\frac{d \lambda}{2\pi} e^{-i\lambda x P^+ }      \langle  h \vert \bar \psi_\beta  (\lambda n) \partial_\perp^\mu  \psi_\alpha (0) \vert h \rangle, 
\end{eqnarray}  
where $D^\mu (x)$ is the covariant derivative $D^\mu (x) =\partial^\mu + i g_s G^\mu (x)$.  In the above, we have suppressed gauge links built with that in Eq.(\ref{GLF})  between operators. 
The three types of twist-3 matrix elements are not independent.  One can derive the relation among them as follows:
\begin{eqnarray} 
M^\mu_{D} (x_1,x_2) = \frac{1}{ x_2-x_1 - i\varepsilon} M^\mu_{F} (x_1,x_2) +  2\pi \delta (x_1-x_2) {\mathcal M}^\mu_{\partial } (x_1). 
\label{MRL}             
\end{eqnarray} 
This relation is for matrix elements defined with the past-pointing gauge link in Eq.(\ref{GLF}). In the case of future-pointing gauge links the relation becomes slightly different. We will come back to the difference later.  

The introduced twist-3 matrix elements are parametrized with twist-3 parton distributions.  With respect to symmetries, the parametrization is
\begin{eqnarray} 
  {\mathcal M}^\mu_{F} (x_1,x_2) &=& \frac{1}{2} \biggr [ T_F (x_1,x_2) \tilde s^\mu_\perp  
     + T_{\Delta} (x_1,x_2) is^\mu_\perp \gamma_5 \biggr ] \gamma^- + \frac{1}{4} \biggr [ \lambda \tilde T_{\Delta } (x_1,x_2) i \gamma_5 \gamma^\mu_\perp            + \tilde T_F  (x_1,x_2) i\gamma^\mu_\perp \biggr ]\gamma^-, 
\nonumber\\
  {\mathcal M}^\mu_{D} (x_1,x_2) &=& \frac{1}{2} \biggr [ D_F (x_1,x_2) \tilde s^\mu_\perp  
     + D_{\Delta} (x_1,x_2) is^\mu_\perp \gamma_5 \biggr ] \gamma^- + \frac{1}{4} \biggr [ \lambda \tilde D_{\Delta } (x_1,x_2) i \gamma_5 \gamma^\mu_\perp \           + \tilde D_F  (x_1,x_2) i\gamma^\mu_\perp \biggr ]\gamma^ -, 
\nonumber\\
  {\mathcal M}^\mu_{\partial } (x ) &=&\frac{1}{2} \biggr[ - i\gamma_5 \gamma^- s_\perp^\mu q_\partial (x) 
     - i \gamma^- \tilde s^\mu q'_\partial (x) \biggr ] + \frac{1}{4}  \biggr[ - i\lambda \gamma_5 \gamma_\perp^\mu  \gamma^-  h_\partial  (x) 
     + \gamma_\perp^\mu \gamma^- e_\partial (x)  \biggr ],              
\label{tw3}             
\end{eqnarray} 
where $\tilde s_\perp^\mu$ is given by $\tilde s_\perp^\mu = \epsilon_\perp^{\mu\nu} s_{\perp\nu}$. 
From Hermiticity and symmetries of parity and time reversal, one can find the properties of twist-3 parton distributions in $M_{D,F}^\mu$,
\begin{eqnarray} 
     && T_F (x_1,x_2) =  T_F(x_2,x_1), \quad \quad  T_{\Delta} (x_1,x_2) = - T_{\Delta} (x_2,x_1), \quad\quad
 \tilde T_F (x_1,x_2) = \tilde T_F (x_2,x_1), 
 \nonumber\\
     &&
 \tilde T_\Delta (x_1,x_2) =- \tilde T_\Delta (x_2,x_1) \quad\quad  D_F (x_1,x_2) =  - D_F(x_2,x_1), \quad \quad  D_{\Delta} (x_1,x_2) = D_{\Delta} (x_2,x_1), 
 \nonumber\\
     && \tilde D_F (x_1,x_2) = -\tilde D_F (x_2,x_1), \quad\quad \tilde D_\Delta (x_1,x_2) = \tilde D_\Delta (x_2,x_1)   .     
\label{TW3P}
\end{eqnarray}
From the relation in Eq.(\ref{MRL}), we have the following relations between twist-3 parton distributions relevant to our work:
\begin{eqnarray} 
   D_\Delta(x_1,x_2)  &=&  \left ( P \frac{1}{x_2-x_1} \right )   T_\Delta  (x_1,x_2) -2\pi \delta(x_1-x_2) q_\partial (x_1),        
\nonumber\\   
  D_F (x_1, x_2) &=&  \left ( P\frac{1}{x_2-x_1} \right )  T_F (x_1,x_2), \quad T_F (x,x) =  + 2 q'_\partial (x), 
\nonumber\\
 \tilde D_\Delta (x_1,x_2) & =&\left ( P\frac{1}{x_2-x_1} \right )  \tilde T_{\Delta} (x_1,x_2)-2\pi \delta (x_2-x_1) h_\partial (x_1),  
 \nonumber\\
    \tilde D_F (x_1,x_2) &=& \left ( P\frac{1}{x_2-x_1} \right )  \tilde T_{F} (x_1,x_2), \quad  \tilde T_F (x,x) 
    = 2  e_\partial (x),    
\end{eqnarray} 
where $P$ stands for the principal-value prescription. As mentioned, these relations are of parton distributions defined with the past-pointing gauge link for Drell-Yan processes. For semi-inclusive deeply inelastic scatterings(SIDIS), one should use 
future-pointing gauge links to define parton distributions. With parity and time reversal symmetries one can show that the twist-2 parton distributions defined with past-pointing gauge links are the same defined with future-pointing gauge links. However, this is not the case for twist-3 parton distributions, especially for those defined with ${\mathcal M}_\partial^\mu$ because of the transverse derivative acting on gauge links. Taking $q_\partial'$ as an example, with parity and time reversal symmetries  one can show that $q_\partial'$, with the future-pointing  gauge link, is $-q_\partial'$ defined with the past-pointing gauge link. Therefore, the second relation in the second line of the above equation is for Drell-Yan processes. For the corresponding relation in SIDIS, the $+$ should be changed into $-$, as noticed in \cite{CMZ}.  It is noted that without parity and time reversal symmetries one already can show that there is a nonzero difference between the two parton distributions defined 
with different gauge links, and the difference is proportional to $T_F(x,x)$\cite{BMPT}.

From equation of motion some relations between 
twist-3 parton distributions defined with quark-gluon correlators and those defined with quark density matrix can be derived\cite{BD,JiG2,EKT,ZYL}. 
There are the following relations 
between twist-3 parton distributions relevant to our work:   
\begin{eqnarray} 
&& \frac{1}{2\pi}\int d x_1  \left ( P\frac{1}{x_2-x_1} \right )   \biggr [ T_F (x_1,x_2)  - T_\Delta (x_1,x_2) \biggr ] = x_2 q_T (x_2) - q_\partial (x_2), 
\nonumber\\
  && \frac{1}{2\pi} \int  d x_1\left ( P\frac{1}{x_2-x_1} \right )  \tilde T_\Delta (x_1,x_2) = h_\partial (x_2) -\frac{1}{2} x_2 h_L (x_2). 
  \label{RL1}
\end{eqnarray}

The process we study is effectively annihilation of quark and antiquark into a virtual photon. The antiquark distributions can be obtained from definitions of quark distributions through a charge-conjugation transformation of the operators in the definitions. We obtain the following relations between antiquark and quark distributions at twist-2: 
\begin{eqnarray} 
 \bar q (x) = - q(-x), \quad \Delta \bar q (x) = \Delta q (-x), \quad \bar h_1 (x) = - h_1 (-x), 
 \label{RL2} 
\end{eqnarray}
and the relations at twist-3,
\begin{eqnarray} 
 &&  \bar e(x) = e(-x), \quad \bar h_L (x) = - h_L(-x), \quad \bar q_T (x) = q_T (-x)
\nonumber\\
  && \bar e_\partial (x)  = e_\partial (-x), \quad \bar h_\partial (x) = h_\partial (-x), \quad \bar q'_\partial (x) = q'_\partial (-x), \quad \bar q_\partial (x) = -q_\partial (-x) . 
 \label{RL3} 
\end{eqnarray}   
In the above, antiquark distributions are in the left-hand side of each equation. In our notation all twist-2 parton distributions are dimensionless. All twist-3 parton distributions have the dimension one in mass.

\par\vskip10pt 

\par 
\begin{figure}[hbt]
\begin{center}
\includegraphics[width=14cm]{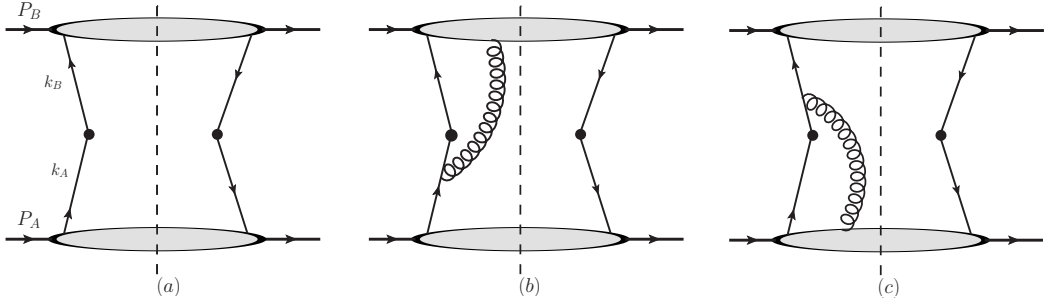}
\end{center}
\caption{Tree-level diagrams for $W^{\mu\nu}$ of Drell-Yan processes. The black dots represent the insertion of the electromagnetic current.   } 
\label{Feynman-dg1}
\end{figure}

\par\vskip10pt
\noindent
{\bf 3. THE DOUBLE-SPIN DEPENDENT PART OF THE HADRONIC TENSOR AND DOUBLE-SPIN ASSYMETRIES} 
\par 

\par\vskip5pt

At the leading order of $\alpha_s$,  the contributions to $W^{\mu\nu}$ are from diagrams  given by Fig.1. 
In these diagrams the upper and lower bubble represent jetlike Green functions related 
to the longitudinally polarized hadron $h_B$ in the initial state,  and  transversely polarized $h_A$,  respectively. The middle part in the diagrams consist of explicit Feynman diagrams of parton scattering. 
In this work we use the Feynman gauge. 
Since the jet functions are jetlike, 
there are power counting for momenta of partons from $h_A$ or $h_B$, respectively. 
For example, in Fig.1(a) the momenta 
$k_A$ and $k_B$ scale like 
\begin{equation} 
 k_A^\mu  \sim Q (1,\lambda^2,\lambda,\lambda), \quad  k_B^\mu \sim Q (\lambda^2, 1,\lambda,\lambda),
\label{PC} 
\end{equation}
and momenta of gluons scale similarly.   The gluon 
field vectors also scale like the pattern of their momentum as in Eq.(\ref{PC}) in the 
gauge we work.

\par 

In collinear factorization one needs to expand the contributions from Fig. 1 
in power of $\lambda$. We first consider diagrams in Fig.1(a).  The contribution 
can be written in the form
\begin{equation} 
W^{\mu\nu} (P_A,P_B,q) \biggr\vert_{1a} = \int d^4 k_A d^4 k_B  \Gamma _{ji} (P_A,k_A) 
H^{\mu\nu}_{ij, lk} (k_A,k_B,q)  \bar  \Gamma_{kl}(P_B,k_B),
\end{equation} 
where $\Gamma_{A,B}$ are quark density matrices represented by the lower and upper bubble, respectively. They are given by 
\begin{eqnarray}
\Gamma_{ji}(P_A,k_A ) &=& \int \frac{ d^4\xi} {(2\pi)^4} e^{-i \xi \cdot k_A}
\langle h_A (P_A) \vert  \left [ \bar q(\xi) \right ]_i
 \left [ q (0)\right ]_j \vert h_A (P_A) \rangle ,
\nonumber\\
 \bar \Gamma_{ij}(P_B,k_B) &=& \int \frac{ d^4\xi} {(2\pi)^4} e^{-i \xi \cdot k_B}
\langle h_B (P_B) \vert  \left [  q(\xi) \right ]_i
 \left [ \bar q (0)\right ]_j \vert h_B (P_B) \rangle ,
\end{eqnarray}
where $ij$ stand for Dirac and color indices. 
Note that $H^{\mu\nu}_{ij,kl} (k_A,k_B,q)$ is the middle part of Fig.1(a)  which is given by
\begin{equation} 
  H^{\mu\nu}_{ij,lk} (k_A,k_B,q) =   \delta^4 ( k_A+k_B -q)  \biggr [ \gamma^\mu \biggr ]_{lj} 
    \biggr [ \gamma^\nu \biggr ]_{ik}.  
\end{equation} 
To find the contributions, we expand $H$ in $\lambda$,
\begin{equation} 
  H^{\mu\nu}_{ij,lk} (k_A,k_B,q) =   \delta^4 ( \hat k_A+ \hat k_B -q)  \biggr [ \gamma^\mu \biggr ]_{lj} 
    \biggr [ \gamma^\nu \biggr ]_{ik} - (k_A + k_B)^\rho_\perp \frac{\partial \delta^4( \hat k_A + \hat k_B-q)}
             {\partial q_\perp^\rho}  \biggr [ \gamma^\mu \biggr ]_{lj} 
    \biggr [ \gamma^\nu \biggr ]_{ik}   + {\mathcal O} (\lambda^2),  
\end{equation} 
with 
\begin{equation} 
   \hat k_A^\mu = (k_A^+, 0,0,0), \quad \hat k_B^\mu  = (0,k_B^-,0,0). 
\end{equation}    
The terms of ${\mathcal O} (\lambda^2)$ will give contributions beyond twist-3 and can safely be neglected.
With the expansion $W^{\mu\nu}$ from Fig.1(a) becomes:
\begin{eqnarray} 
W^{\mu\nu} (P_A,P_B,q) \biggr\vert_{1a} &=&  \int d k_A^+  d k_B^-  
 \biggr [ \gamma^\mu \biggr ]_{lj}  \biggr [ \gamma^\nu \biggr ]_{ik} \int \frac{ d\xi^- } {2\pi } e^{-i \xi^-  k_A^+ }  \int \frac{ d \xi^+  } {2\pi} e^{-i \xi^+  k_B^- } \biggr \{ \delta^4 ( \hat k_A + \hat k_B  - q) \nonumber\\
    && 
\langle h_A (P_A) \vert  \left [ \bar q(\xi^- n ) \right ]_i
 \left [ q (0)\right ]_j \vert h_A (P_A) \rangle  
\langle h_B (P_B) \vert  \left [  q(\xi^+ l ) \right ]_k
 \left [ \bar q (0)\right ]_l \vert h_B (P_B) \rangle  
\nonumber\\
   &&  +  i \frac{\partial \delta^4( \hat k_A + \hat k_B-q)}
             {\partial q_\perp^\rho}
 \biggr [ \langle h_A (P_A) \vert  \left [ \partial_{\perp\rho} \bar q(\xi^- n ) \right ]_i
 \left [ q (0)\right ]_j \vert h_A (P_A) \rangle  
\nonumber\\
&& 
\langle h_B (P_B) \vert  \left [  q(\xi^+ l ) \right ]_k
 \left [ \bar q (0)\right ]_l \vert h_B (P_B) \rangle +
\langle h_A (P_A) \vert  \left [  \bar q(\xi^- n ) \right ]_i
 \left [ q (0)\right ]_j \vert h_A (P_A) \rangle 
\nonumber\\
    &&  
\langle h_B (P_B) \vert  \left [  \partial_{\perp\rho} q(\xi^+ l ) \right ]_k
 \left [ \bar q (0)\right ]_l \vert h_B (P_B) \rangle 
   \biggr ]  \biggr \}  + \cdots, 
\label{CEXP}     
\end{eqnarray} 
where $\cdots$ stands for terms beyond twist-3. Using the parametrizations of matrix elements discussed in the last section,  we obtain the following:
\begin{eqnarray} 
W^{\mu\nu} \biggr\vert_{1a} 
   &=& \frac{1}{2 N_c} \lambda  \biggr \{  \frac{\partial \delta^2 (q_\perp)}{\partial q_\perp^\rho} \biggr [  \bar h_\partial (y) h_1 (x) \biggr ( g_{\perp}^{\rho\mu} s_\perp^\nu + g_\perp^{\rho\nu} s_\perp^\mu - g_\perp^{\mu\nu} s_\perp^\rho \biggr ) - 2 \Delta \bar q(y) q_\partial (x) s_\perp^\rho g_\perp^{\mu\nu} \biggr ] 
\nonumber\\
   && + \delta^2 (q_\perp) \frac{1}{P_A\cdot P_B} \biggr [ -2 \Delta \bar q(y) q_T (x)  ( P_B^\mu s_\perp^\nu +    
P_B^\nu s_\perp^\mu ) + \bar h_L (y) h_1 ( x)  ( P_A^\mu s_\perp^\nu +    
P_A^\nu s_\perp^\mu ) \biggr ] \biggr\} , 
\label{F1a} 
\end{eqnarray}   
where $\lambda$ is the helicity of $h_B$ and $s_\perp^\mu$ is the transverse spin vector of $h_A$. 
$q^+$, and $q^-$ are given by
\begin{equation} 
   q^+ = x P_A^+, \quad q^- = y P_{B}^-. 
 \label{XY} 
 \end{equation} 
Here, and in the below, we always use the notation that parton distributions with the variable $x$ or $y$ are of $h_A$ or of $h_B$, respectively. 
It is noted that the above results are not exactly only from Fig.1(a) because of that the parton distributions contain gauge links. To find the contributions of gauge links, we need to consider diagrams of one gluon 
exchange as Fig.1(b) and Fig,1(c) and diagrams with exchanges of more gluons. 
We will call the gluons with the polarization index $-$ or $+$ as $G^-$ or $G^+$ gluons, respectively.             
With the power counting discussed around Eq.(\ref{PC}), one easily finds that the leading contributions from the exchange of any number of 
$G^-$ gluons with the upper bubble or $G^+$ gluons with the lower bubble are at the same leading power of $\lambda$ as that of Fig.1(a). The summation of these contributions can be done with Ward identity in a standard way. The summation gives the contributions of gauge links in the parton distributions.

\par 
Figures 1(b) and 1(c) give the so-called three-parton contributions, where the exchanged gluon can be transversely polarized. In Fig.1(b) the twist-3 contribution involves twist-2 quark  distribution $h_1$ of $h_A$ and twist-3 parton distributions of $h_B$. The calculation is straightforward.  
It is found that the contribution from the transversely polarized gluon and $G^-$ gluon can be summed into the form which involves the field strength tensor operator $G^{-\mu}$.  We obtain the contribution of Fig.1(b) and its complex conjugates as follows:
\begin{equation} 
W^{\mu\nu}\biggr\vert_{1b + c.c.} = \frac{\lambda_B }{2\pi N_c} \delta^2 ( q_\perp) \frac{1}{x P_A\cdot P_B} ( P_B^\mu s_\perp^\nu + P_B^\nu s_\perp^\mu ) h_1 (x) \int \frac{ d y_1}{y-y_1}  \tilde T_{\Delta} (-y, -y_1 ).
\label{F1b} 
\end{equation} 
Similarly, we obtain the contribution from Fig.1(c) and its complex conjugates as follows:
\begin{equation} 
W^{\mu\nu}\biggr\vert_{1c + c.c.} = - \frac{\lambda_B }{2\pi N_c} \delta^2 ( q_\perp) \frac{1}{y P_A\cdot P_B} ( P_A^\mu s_\perp^\nu + P_A^\nu s_\perp^\mu )  \Delta \bar q (y)  \int \frac{ d x_1}{x-x_1} 
    ( T_\Delta (x_1,x) - T_F (x_1,x)). 
\label{F1c}
\end{equation} 
Again, these contributions, in fact, contain contributions of diagrams with exchanges of more than one $G^+$ or $G^-$ gluon. These contributions can be summed into the form of gauge links in parton distributions as discussed after Eq.(\ref{XY}). 
It is interesting to note that the involved integrals of twist-3 parton distributions in the above can be expressed with two-parton distributions by using the relation in Eq.(\ref{RL1}) and the relations between quark and antiquark distributions in Eqs.(\ref{RL2}) and (\ref{RL3}).

\par
 The results in Eqs.(\ref{F1a}), (\ref{F1b}), and (\ref{F1c}) are the contributions for the partonic process, in which an initial quark or antiquark comes from $h_A$ or $h_B$, respectively. 
 The complete hadronic tensor is the sum of the contributions in Eqs.(\ref{F1a}), (\ref{F1b}), and (\ref{F1c}) and 
 the contributions from the charge-conjugated partonic process. 
 The sum is
\begin{eqnarray}
W^{\mu\nu} &=& \frac{\delta^2 (q_\perp) \lambda} { N_c P_A\cdot P_B} \biggr \{ 
    \biggr ( P_B^\mu s_\perp^\nu + P_B^\nu s_\perp^\mu \biggr ) \biggr [ -  q_T (x) \Delta \bar q(y)  +\frac{1}{x}   h_1 (x) (  \bar h_\partial (y) -\frac{1}{2}  y \bar h_L (y) \biggr ] 
\nonumber\\    
    && 
 + \biggr ( P_A^\mu s_\perp^\nu + P_A^\nu s_\perp^\mu \biggr )  \biggr [ - \frac{1}{y}  ( q_\partial  (x) - x q_T (x)) \Delta \bar q(y)    + \frac{1}{2} h_1 (x)  \bar h_L (y)   \biggr ] \biggr \} 
\nonumber\\
   && + \frac{\partial \delta^2 (q_\perp) } {\partial q_\perp^\rho} \frac{\lambda}{N_c} \biggr \{ 
       ( g_\perp^{\rho\mu} s_\perp^\nu +  g_\perp^{\rho\nu} s_\perp^\mu - g_\perp^{\nu\mu} s_\perp^\rho ) 
          \frac{1}{2}   h_1 (x)  \bar h_\partial (y)  - s_\perp^\rho g_\perp^{\mu\nu}    q_\partial (x)  \Delta \bar q(y)  \biggr \}  + (q \leftrightarrow \bar q),  
\label{WMN}                                
\end{eqnarray} 
where the notation $(q \leftrightarrow \bar q)$ stands for the contribution of the charge-conjugated partonic process. It is obtained by replacing the combination  $a(x) \bar b(y)$ of parton distributions with  $\bar a(x) b(y)$, where $a(x)$ or $\bar b (y)$ is a parton distribution of $h_A $ or $h_B$, respectively.  
This result contains $\delta^2(q_\perp)$ and its derivative. Therefore, one should take the result as a distribution of $q_\perp$. The $U_{em} (1)$ gauge invariance should be then understood as
\begin{equation} 
   \int d^2 q_\perp {\mathcal F}(q_\perp) q_\mu W^{\mu\nu} = \int d^2 q_\perp  {\mathcal F}(q_\perp) q_\nu W^{\mu\nu} =0, 
 \end{equation} 
 where ${\mathcal F}(q_\perp)$ is a test function. 
Our result satisfies this equation and hence is gauge invariant.  Because the hadronic tensor at the order of $\alpha_s$ is a distribution of $q_\perp$, in predictions of relevant physical observables,  $q_\perp$ is integrated over. 
With this in mind, the introduced weight observable in Eq.(\ref{WDSIG}) with ${\mathcal F}=1$ will be 
determined by the part with $\delta^2 (q_\perp)$ in $W^{\mu\nu}$, the observables with ${\mathcal F}$ proportional to $q_\perp$ will be determined by the part with the derivative of $\delta^2 (q_\perp)$ in $W^{\mu\nu}$.      
 
Before ending this section, it is worth discussing the physical meaning of terms with the derivative of $\delta^2(q_\perp)$ in the hadronic tensor. At the considered order $\alpha_s^0$, the quark and the antiquark, which annihilate into the virtual photon, are partons directly from initial hadrons. They have only intrinsic nonzero but small transverse momenta at order of $\Lambda_{QCD}$. At the leading power of the inverse of $Q$ or at leading twist,   these momenta are neglected. It results in that the hadronic tensor at the order is proportional to $\delta^2(q_\perp)$. At the next-to-leading power, the effect of the nonzero, but small transverse momenta, has to be taken into account. This effect is included, e.g., in the second term in Eq.(\ref{CEXP}), 
which gives the contributions to $W^{\mu\nu}$ proportional to the derivative of $\delta^2(q_\perp)$. It is also noted that beyond 
the leading order $\alpha_s$, the annihilated quark and antiquark can have large transverse momenta because of hard gluon radiations.  For our observables in the next section, the effects of hard gluon radiations are suppressed by $\alpha_s$.

\par\vskip20pt
\noindent 
{\bf 4. PHYSICAL OBSERVABLES}

In this section we consider experimental observables related to the polarizations of initial hadrons. 
We consider the angular distribution of the final lepton in the Collins-Soper frame\cite{CS-frame}.  In this frame 
the momentum $k_1$ of the lepton is 
\begin{equation}  
   k_1^\mu = \frac{Q}{2} ( 1, \sin\theta \cos\phi, \sin\theta\sin\phi, \cos\theta),
\end{equation}     
where $\theta$ is the polar angle between the lepton momentum and $Z$ axis, which bisects 
the angle between directions of the two initial hadrons. Note that $\phi$ is the azimuthal angle between $k_{1\perp}$ 
and $q_\perp$.  It is noted that the transverse spin vector in the laboratory frame given in Sect.II is not exactly the same in the Collins-Soper frame.  In the considered case of small $q_\perp$, the difference is at order 
of $q_\perp^2$, which can be safely ignored. We denote the azimuthal angle between $s_\perp$ and $q_\perp$ as $\phi_s$. 

The angular distribution can be derived by introducing four covariant vectors as coordinate vectors as discussed in \cite{BQR}.  With our hadronic tensor, the contribution proportional to $\lambda \vert s_\perp \vert$ is obtained as follows:
\begin{equation} 
  \frac{d\sigma}{d Q^2 d\Omega} = \lambda \vert s_\perp\vert  \sum_q\frac{(4 \pi \alpha)^2 e_q^2}{64\pi^2   N_c Q^3} \int d x dy  \delta (x y S -Q^2) \biggr [  x W_1 - y W_2 \biggr ] \sin(2\theta) \cos(\Delta\phi_{s\ell} ) , 
\label{dsigma}   
\end{equation}      
where $\Delta\phi_{s\ell}$ is the difference $\phi-\phi_s$. Note that $W_{1,2}$ is given by the first line and second line 
of $W^{\mu\nu}$ in Eq.(\ref{WMN}), 
\begin{eqnarray} 
W_1 &=&  -q_T (x) \Delta \bar q(y) + \frac{1}{ x} h_1 (x) \left [  \bar h_\partial (y) 
   -  \frac{1}{2} y \bar h_L (y) \right ]  + (q\leftrightarrow \bar q), 
\nonumber\\
W_2 &=& - \frac{1}{y} \Delta \bar q (y) \left [ q_\partial (x) - x q_T(x) \right ]  + \frac{1}{2} \bar h_L (y) h_1 (x) + (q\leftrightarrow \bar q) . 
\end{eqnarray}     
This result is in agreement with that in \cite{BMT}. 
The differential cross section given in Eq.(\ref{dsigma}) is determined by the nonderivative part of $W^{\mu\nu}$. To see the effect of the derivative part, one has to consider the weighted differential cross section introduced in Eq.(\ref{WDSIG}). It is noted that in \cite{BMT}
the result of the hadronic tensor with collinear parton distributions is given where $q_\perp$ is integrated over. After the integration, only the nonderivative part remains, and the derivative part gives no contribution 
to the result. Therefore, with the result in \cite{BMT} one cannot make predictions of weighted differential cross sections with weights involving $q_\perp$.   
We introduce two weights, 
\begin{equation} 
   {\mathcal F}_1 = \frac{1}{Q} q_\perp\cdot s_\perp, \quad   {\mathcal F}_2 = \frac{1}{Q} q_\perp\cdot \tilde s_\perp. 
\end{equation} 
The corresponding weighted differential cross sections are: 
\begin{eqnarray} 
 \frac{d\sigma \langle {\mathcal F}_1\rangle }{d Q^2 d\Omega}  &=&  - \lambda \vert s_\perp \vert^2\sum_q  \frac{ (4 \pi \alpha)^2 e_q^2}{64\pi^2 N_c Q^3 } \int d x dy  \delta (x y S -Q^2)  \biggr [ D_1 (1+\cos^2\theta ) 
   - D_2 \sin^2 \theta \cos (2\Delta \phi_{s\ell}) \biggr ], 
\nonumber\\
 \frac{d\sigma \langle {\mathcal F}_2 \rangle }{d Q^2 d\Omega}  &=&  - \lambda \vert s_\perp \vert^2 \sum_q\frac{ (4 \pi \alpha)^2 e_q^2}{64 \pi^2 N_c Q^3} \int d x dy  \delta (x y S -Q^2)  \biggr [ D_2 \sin^2\theta \sin(2 \Delta\phi_{s\ell} ) \biggr ] , 
 \end{eqnarray} 
 where $D_{1,2}$ are given by the derivative part of $W^{\mu\nu}$, 
 \begin{eqnarray} 
 D_1 = - \Delta \bar q (y) q_\partial (x) -  \Delta q (y)  \bar q_\partial (x), 
 \quad 
 D_2 =\frac{1}{2 } (\bar h_\partial (y) h_1 (x) +  h_\partial (y) \bar h_1 (x)). 
\end{eqnarray}     
From these spin-dependent differential cross sections, one can define various double-spin asymmetries by taking their differences between those with $\lambda = \pm 1$. It is interesting to note that only the differential 
cross section weighted with ${\mathcal F}_1$ can be measured if the azimuthal angle or the solid angle is integrated. In this case we have only one observable remaining nonzero,
\begin{equation} 
\frac{d\sigma \langle {\mathcal F}_1\rangle }{ d Q^2 }  =  - \lambda \vert s_\perp \vert^2 \sum_q\frac{ (4 \pi \alpha )^2 e_q^2}{12\pi N_c  Q^3 } \int d x dy  \delta (x y S -Q^2)   D_1 .   
\end{equation} 
With this result, a simple asymmetry can be defined as
\begin{eqnarray} 
 A\langle {\mathcal F}_1\rangle_{LT}  = 
 \frac{\frac{ d\sigma \langle {\mathcal F}_1\rangle }{ d x d Q^2}  (\lambda=1, s_\perp) - \frac{ d\sigma \langle {\mathcal F}_1\rangle }{ d x d Q^2}  (\lambda=-1, s_\perp)}{\frac{d\sigma_0}{dx d Q^2}  (\lambda=1, s_\perp) +\frac{d\sigma_0}{dx d Q^2}   (\lambda=-1, s_\perp)} 
=  \frac{1}{Q} \vert s_\perp\vert ^2 
   \frac{  \sum_q  e_q^2 (\bar q_\partial (x) \Delta q (y) +  q_\partial (x)  \Delta \bar q (y))}{ \sum_q e_q^2 ( q(x) \bar q(y) 
     + \bar q(x)  q(y) )} , 
\end{eqnarray} 
where $\sigma_0$ is the unpolarized cross section at the leading power. In the above $y$ is fixed as $Q^2/x S$.  Similar asymmetries in angular distributions can also be constructed with the results here. 

\par\vskip20pt
\noindent
{\bf 5. SUMMARY}

We have made an analysis for double-spin asymmetries in Drell-Yan processes.  The asymmetries arise at twist-3 level. The complete part of the hadronic tensor relevant to these asymmetries is derived. 
This twist-3 part contains not only a contribution 
of $\delta^2(q_\perp)$ as twist-2 parts do, but also a contribution proportional to the derivative of the $\delta$ function.  Only the sum of the two contributions as a distribution of $q_\perp$ is gauge invariant.  Based on our results, observables are constructed to identify the spin effects. From these observables one can build double-spin asymmetries, which can be used for extracting twist-3 parton distributions.    

\par\vskip40pt

\noindent
{\bf Acknowledgments}
\par
The work is supported by National Natural Science Foundation of P.R. China(Grants No. 12075299,11
821505, 11847612,11935017 and 12065024)  and by the Strategic Priority Research Program of Chinese Academy of Sciences Grant No. XDB34000000.

\par

\end{document}